\documentclass[sigconf, nonacm]{acmart}
\usepackage{xcolor}
\usepackage{amsmath}
\usepackage{graphicx}
\usepackage{textcomp}
\usepackage{algpseudocode}
\usepackage{algorithm}
\usepackage{enumitem}
\usepackage{subcaption}
\usepackage{bm}
\usepackage{lipsum}
\usepackage{caption}
\usepackage{listings}
\usepackage{calc}
\usepackage{amsfonts}
\usepackage{tabu}
\usepackage{tikz}
\usepackage{makecell}
\usetikzlibrary{arrows.meta}
\usepackage{pgfplots}
\usepackage{multirow}
\usepackage{flushend}
\usepackage{booktabs}
\usepackage{hyperref}
\usepackage{cleveref}
\usepackage{soul}

\AtBeginDocument{%
  \providecommand\BibTeX{{%
    \normalfont B\kern-0.5em{\scshape i\kern-0.25em b}\kern-0.8em\TeX}}}

\definecolor{codegreen}{rgb}{0,0.6,0}
\definecolor{codegray}{rgb}{0.5,0.5,0.5}
\definecolor{codepurple}{rgb}{0.58,0,0.82}
\definecolor{backcolour}{rgb}{0.95,0.95,0.92}
\definecolor{jcred}{HTML}{e31a1c}
\definecolor{jcgreen}{HTML}{33a02c}
\definecolor{jcblue}{HTML}{1f78b4}
\definecolor{jcorange}{HTML}{ff7f00}
\definecolor{jcpurple}{HTML}{6a3d9a}
\definecolor{jclightred}{HTML}{fb8072}
\definecolor{jclightgreen}{HTML}{b3de69}
\definecolor{jclightblue}{HTML}{80b1d3}
\definecolor{jclightorange}{HTML}{fdb462}
\definecolor{jclightpurple}{HTML}{bebada}
\definecolor{jcredl}{HTML}{fb8072}
\definecolor{jcgreenl}{HTML}{b3de69}
\definecolor{jcbluel}{HTML}{80b1d3}
\definecolor{jcorangel}{HTML}{fdb462}
\definecolor{jcpurplel}{HTML}{bebada}
\definecolor{jcbluem}{HTML}{488bb8}

\lstdefinestyle{mystyle}{
  frame=tblr,
  commentstyle=\color{codegreen},
  keywordstyle=\color{codepurple},
  basicstyle=\scriptsize\ttfamily,
  breakatwhitespace=false,         
  breaklines=true,                 
  captionpos=b,                    
  keepspaces=true,                 
  numbers=left,                    
  numbersep=5pt,                  
  showspaces=false,                
  showstringspaces=false,
  showtabs=false,                  
  tabsize=2,
  escapeinside={(*@}{@*)},
}
\lstdefinelanguage{maseir}{%
  language     = python,
  morekeywords = {in, return, rate, MXInt},
}

\definecolor{jcblue}{HTML}{1f78b4}

\newcommand*\jc[1]{\textcolor{jcblue}{\bf JC: #1}}

\newcommand*\best[1]{\textcolor{jcgreen}{\bf #1}}

\newcommand\varied[1]{\textcolor{jcred}{\boldsymbol{#1}}}

\usepackage{pifont}

\newcommand*\change[1]{\textcolor{blue}{#1}}

\newcommand*\sw[1]{\textcolor{jcblue}{\tt #1}}
\newcommand*\hw[1]{\textcolor{jcred}{\tt #1}}

\begin{document}

\title{A Dataflow Compiler for Efficient LLM Inference using Custom Microscaling Formats}

\author{Jianyi Cheng}
\affiliation{
\institution{University of Cambridge, UK}
\country{}
}
\email{jianyi.cheng@cl.cam.ac.uk}

\author{Cheng Zhang}
\affiliation{\institution{Imperial College London, UK}
\country{}
}
\email{cheng.zhang122@imperial.ac.uk}

\author{Zhewen Yu}
\affiliation{\institution{Imperial College London, UK}
\country{}
}
\email{zhewen.yu18@imperial.ac.uk}

\author{Christos-Savvas Bouganis}
\affiliation{\institution{Imperial College London, UK}
\country{}
}
\email{christos-savvas.bouganis@imperial.ac.uk}

\author{George A. Constantinides}
\affiliation{\institution{Imperial College London, UK}
\country{}
}
\email{g.constantinides@imperial.ac.uk}

\author{Yiren Zhao}
\affiliation{\institution{Imperial College London, UK}
\country{}
}
\email{a.zhao@imperial.ac.uk}

\renewcommand{\shortauthors}{Cheng, et al.}

\begin{abstract}
Model quantization represents both parameters (weights) and intermediate values (activations) in a more compact format, thereby directly reducing both computational and memory cost in hardware. The quantization of recent large language models (LLMs) faces challenges to achieve competitive memory density compared to other models such as convolutional neural networks, since values in LLMs require larger dynamic ranges.

Current hardware can expedite computation for LLMs using compact numerical formats such as low-bitwidth integers or floating-point numbers. Each has advantages: integer operations simplify circuit design, whereas floating-point calculations can enhance accuracy when a wider dynamic range is required. In this work, we seek an efficient data format that combines the best of both worlds: Microscaling (MX) formats. MX formats are efficient data formats that achieve both large dynamic ranges and high memory density.

In this paper, we propose a compiler named MASE for exploring mixed-precision MX formats on dataflow hardware accelerators for LLM inference. Our main contributions are twofold. First, we propose a novel orchestration abstraction to explore both software and hardware optimizations with new data formats. Second, MASE achieves LLM inference at an average precision of 4-bits, with minimal to no accuracy degradation. To our knowledge, MASE represents the first effort to harness fine-grain multi-precision MX formats in the design of LLM hardware accelerators. Over a range of LLMs and datasets, MASE achieves an average improvement of 24\% in $\Delta$ accuracy with an overhead of only 3\% in energy efficiency compared to designs using 8-bit fixed-point numbers.
\end{abstract}


\maketitle

\section{Introduction}

Large Language Models (LLMs)~\cite{brown2020language,gpt-neo,zhang2022opt, touvron2023llama, chiang2023vicuna, touvron2023llama2} have gained significant attention, with empirical evidence suggesting that models must reach a certain scale to exhibit \textit{emergent abilities} \cite{wei2022emergent}. These large models, such as {\tt GPT-3}, {\tt Vicuna} and {\tt LLaMA}, are pre-trained on vast amounts of text data, enabling them to provide state-of-the-art results in areas like language translation~\cite{feng2020language}, question-answering~\cite{yang2020bert}, sentiment analysis~\cite{liu2021makes}. One of the main challenges in LLM inference is the vast number of parameters involved \cite{zhang2022full}. For example, the larger variants in the {\tt GPT} family can have hundreds of billions of parameters, which would require a minimum of 300 GB of memory to store them in a {\tt FP16} format \cite{brown2020language}. To reduce memory size, quantization is employed to reduce the precision of both model parameters and activations to a more compact representation.

\begin{table}
\centering
\caption{Evaluation of MX formats in similar average bits for quantizing {\tt LLaMA} on {\tt Wikitext2}. Small perplexity means better LLM performance. Higher memory density or hardware arithmetic density (both defined by Darvish \textit{et al.}~\cite{darvish2020pushing}) means better hardware efficiency.}
\label{tab:motivation}
{
\begin{tabular}{ccrrr}
\toprule
Approaches & Config & \multicolumn{1}{c}{Perplexity} & \makecell[c]{Memory \\ Density} & \makecell[c]{Arithmetic \\ Density} \\
\midrule
{\tt FP32} & - & \best{7.06} & 1$\times$ & 1$\times$ \\
\tt{Int8} & W8A8 & 265 & 4$\times$ & 7.7$\times$ \\
\tt{FP8} & W8A8 & 7.18 & 4$\times$ & 17.4$\times$ \\
\midrule
\tt{MXInt8} & W8A8 & \best{7.07} & 3.8$\times$ & 14.4$\times$ \\
\tt{BMF8} & W8A8 & 223k & 3.8$\times$ & 14.4$\times$ \\
\tt{BL8} & W8A8 & 18.8 & 3.8$\times$ & 16.1$\times$ \\
\bottomrule
\end{tabular}}
\end{table}
\begin{figure*}
    \centering
\begin{subfigure}[b]{0.18\textwidth}
\definecolor{c1}{HTML}{1b9e77}
\definecolor{c2}{HTML}{d95f02}
\definecolor{c3}{HTML}{7570b3}
\definecolor{c4}{HTML}{e7298a}
\definecolor{c5}{HTML}{a1d99b}
\definecolor{c6}{HTML}{e6ab02}
\definecolor{c7}{HTML}{d7301f}
\definecolor{c8}{HTML}{b15928}
\definecolor{c9}{HTML}{bc80bd}
\definecolor{c10}{HTML}{fdb462}
\begin{tikzpicture}[thick,scale=0.75, every node/.style={scale=0.75}]
\pgfplotsset{compat=1.3}
\begin{axis}[
    height=105mm,
    width=1.5\textwidth,
    xlabel={Block ID},
    ylabel={Variance},
    ylabel style = {align=center},
    xmin=0, xmax=32, 
    ymode=log, 
    log basis y={10},
    ylabel shift = -5 pt,
    scaled x ticks = false, 
    scaled y ticks = false,
    legend columns = 2,
    legend style={at={(0.19,0.26)},anchor=north west, draw=none, fill=none},
    ]
\addplot[dashed, c1, draw=c1, mark=o, line width=1pt, mark options={solid, scale=0.7}] table
[x=block_id, y=data, col sep=space] {q_proj.dat};
\addplot[dashed, c2, draw=c2, mark=o, line width=1pt, mark options={solid, scale=0.7}] table
[x=block_id, y=data, col sep=space] {k_proj.dat};
\addplot[dashed, c3, draw=c3, mark=o, line width=1pt, mark options={solid, scale=0.7}] table
[x=block_id, y=data, col sep=space] {v_proj.dat};

\addplot[dashed, c8, draw=c8, mark=o, line width=1pt, mark options={solid, scale=0.7}] table
[x=block_id, y=data, col sep=space] {mm_a.dat};
\addplot[dashed, c7, draw=c7, mark=o, line width=1pt, mark options={solid, scale=0.7}] table
[x=block_id, y=data, col sep=space] {mm_b.dat};

\addplot[dashed, c6, draw=c6, mark=o, line width=1pt, mark options={solid, scale=0.7}] table
[x=block_id, y=data, col sep=space] {o_proj.dat};

\addplot[dashed, c10, draw=c10, mark=o, line width=1pt, mark options={solid, scale=0.7}] table
[x=block_id, y=data, col sep=space] {up_proj.dat};
\addplot[dashed, c5, draw=c5, mark=o, line width=1pt, mark options={solid, scale=0.7}] table
[x=block_id, y=data, col sep=space] {gate_proj.dat};
\addplot[dashed, c4, draw=c4, mark=o, line width=1pt, mark options={solid, scale=0.7}] table
[x=block_id, y=data, col sep=space] {down_proj.dat};

\addplot[dashed, c9, draw=c9, mark=o, line width=1pt, mark options={solid, scale=0.7}] table
[x=block_id, y=data, col sep=space] {last.dat};

\addlegendentry{$\varied{Q}$}
\addlegendentry{$\varied{K}$}
\addlegendentry{$\varied{V}$}

\addlegendentry{$\varied{A}$}
\addlegendentry{$\varied{B}$}

\addlegendentry{$\varied{B_O}$}

\addlegendentry{$\varied{U}$}
\addlegendentry{$\varied{G}$}
\addlegendentry{$\varied{D}$}

\addlegendentry{$\varied{O}$}

\end{axis}
\end{tikzpicture}
\caption{\footnotesize The variations of the same activation in an LLM have {\em distinct distributions} cross different layers.}
\label{fig:motivation:dist}
\end{subfigure}
\begin{subfigure}[b]{0.15\textwidth}
\centering
\begin{tikzpicture}[thick,scale=0.75, every node/.style={scale=0.75}]
\pgfplotsset{every x tick label/.append style={font=\footnotesize}, compat=1.3}
\begin{axis}[
width=1.5\textwidth,
xbar stacked,
height=100mm,
legend style={at={(1.0,1.0)},anchor=south east, draw=none},
xmin=0, xmax=1,xlabel shift = -2 pt,
xtick={0,0.5,1},
xticklabels={0,0.5, 1},
xlabel={Distribution},
symbolic y coords={%
wqi,
qi,
wki,
ki,
wvi,
vi,  
ai,  
bi,  
w0, 
b0, 
wg, 
g, 
wu, 
u, 
wd,
d,
o
},
yticklabels={      
$W_{Q}$,     
$Q$,      
$W_{K}$,      
$K$,
$W_{V}$,        
$V$,      
$A$,
$B$,      
$W_0$,      
$B_0$,          
$W_G$,      
$G$,      
$W_U$,      
$U$,  
$W_D$,      
$D$,  
$O$
},
ytick=data,
legend columns = 2,
bar width=8,
]

\addplot[jclightgreen, draw=jclightgreen, fill=jclightgreen] table
[y=name, x=3bits, col sep=space] {
name	3bits	4bits	5bits 6bits
wqi	0.25	0.25	0.1666666667	0.3333333333
qi	0.25	0.25	0.3333333333	0.1666666667
wki	0.6666666667	0.08333333333	0.1666666667	0.08333333333
ki	0.25	0.3333333333	0.25	0.1666666667
wvi	0.25	0.25	0.1666666667	0.3333333333
vi	0.1666666667	0.1666666667	0.4166666667	0.25
ai	0.25	0.1666666667	0.3333333333	0.25
bi	0.1666666667	0.25	0.25	0.3333333333
w0	0.25	0.08333333333	0.3333333333	0.3333333333
b0	0.1666666667	0.25	0.3333333333	0.25
wg	0.3333333333	0.25	0.1666666667	0.25
g	0.08333333333	0.3333333333	0.5	0.08333333333
wu	0.25	0.3333333333	0.25	0.1666666667
u	0.25	0.1666666667	0.1666666667	0.4166666667
wd	0.4166666667	0.4166666667	0	0.1666666667
d	0.1666666667	0.08333333333	0.5833333333	0.1666666667
o   0       0       0.09090909091  0.90909090909        
};

\addplot[jclightblue, draw=jclightblue, fill=jclightblue] table
[y=name, x=4bits, col sep=space] {
name	3bits	4bits	5bits 6bits
wqi	0.25	0.25	0.1666666667	0.3333333333
qi	0.25	0.25	0.3333333333	0.1666666667
wki	0.6666666667	0.08333333333	0.1666666667	0.08333333333
ki	0.25	0.3333333333	0.25	0.1666666667
wvi	0.25	0.25	0.1666666667	0.3333333333
vi	0.1666666667	0.1666666667	0.4166666667	0.25
ai	0.25	0.1666666667	0.3333333333	0.25
bi	0.1666666667	0.25	0.25	0.3333333333
w0	0.25	0.08333333333	0.3333333333	0.3333333333
b0	0.1666666667	0.25	0.3333333333	0.25
wg	0.3333333333	0.25	0.1666666667	0.25
g	0.08333333333	0.3333333333	0.5	0.08333333333
wu	0.25	0.3333333333	0.25	0.1666666667
u	0.25	0.1666666667	0.1666666667	0.4166666667
wd	0.4166666667	0.4166666667	0	0.1666666667
d	0.1666666667	0.08333333333	0.5833333333	0.1666666667
o   0       0       0.09090909091  0.90909090909    
};

\addplot[jclightorange, draw=jclightorange, fill=jclightorange] table
[y=name, x=5bits, col sep=space] {
name	3bits	4bits	5bits 6bits
wqi	0.25	0.25	0.1666666667	0.3333333333
qi	0.25	0.25	0.3333333333	0.1666666667
wki	0.6666666667	0.08333333333	0.1666666667	0.08333333333
ki	0.25	0.3333333333	0.25	0.1666666667
wvi	0.25	0.25	0.1666666667	0.3333333333
vi	0.1666666667	0.1666666667	0.4166666667	0.25
ai	0.25	0.1666666667	0.3333333333	0.25
bi	0.1666666667	0.25	0.25	0.3333333333
w0	0.25	0.08333333333	0.3333333333	0.3333333333
b0	0.1666666667	0.25	0.3333333333	0.25
wg	0.3333333333	0.25	0.1666666667	0.25
g	0.08333333333	0.3333333333	0.5	0.08333333333
wu	0.25	0.3333333333	0.25	0.1666666667
u	0.25	0.1666666667	0.1666666667	0.4166666667
wd	0.4166666667	0.4166666667	0	0.1666666667
d	0.1666666667	0.08333333333	0.5833333333	0.1666666667
o   0       0       0.09090909091  0.90909090909    
};

\addplot[jclightred, draw=jclightred, fill=jclightred] table
[y=name, x=6bits, col sep=space] {
name	3bits	4bits	5bits 6bits
wqi	0.25	0.25	0.1666666667	0.3333333333
qi	0.25	0.25	0.3333333333	0.1666666667
wki	0.6666666667	0.08333333333	0.1666666667	0.08333333333
ki	0.25	0.3333333333	0.25	0.1666666667
wvi	0.25	0.25	0.1666666667	0.3333333333
vi	0.1666666667	0.1666666667	0.4166666667	0.25
ai	0.25	0.1666666667	0.3333333333	0.25
bi	0.1666666667	0.25	0.25	0.3333333333
w0	0.25	0.08333333333	0.3333333333	0.3333333333
b0	0.1666666667	0.25	0.3333333333	0.25
wg	0.3333333333	0.25	0.1666666667	0.25
g	0.08333333333	0.3333333333	0.5	0.08333333333
wu	0.25	0.3333333333	0.25	0.1666666667
u	0.25	0.1666666667	0.1666666667	0.4166666667
wd	0.4166666667	0.4166666667	0	0.1666666667
d	0.1666666667	0.08333333333	0.5833333333	0.1666666667
o   0       0       0.09090909091  0.90909090909    
};

\legend{$\leq$ 3 bits, 4 bits, 5 bits, $\geq$6 bits}

\end{axis}
\end{tikzpicture}
\caption{\footnotesize Tensor bit widths across blocks when quantized to mixed-precision MXInt.}
\label{fig:motivation:prec}
\end{subfigure}
\begin{subfigure}[b]{0.15\textwidth}
    \centering
    \includegraphics[width=\textwidth]{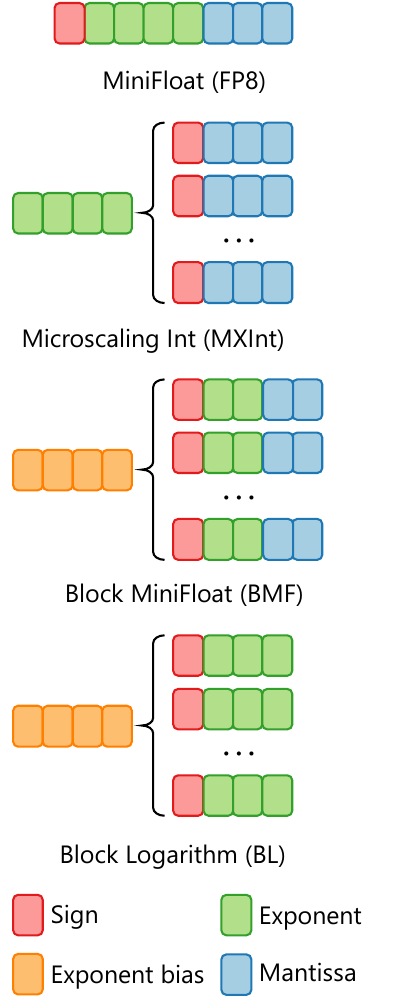} 
    \caption{\footnotesize Examples of FP8 and MX formats.}
    \label{fig:motivation:types}
\end{subfigure}
\begin{subfigure}[b]{0.28\textwidth}
    \includegraphics[width=\textwidth]{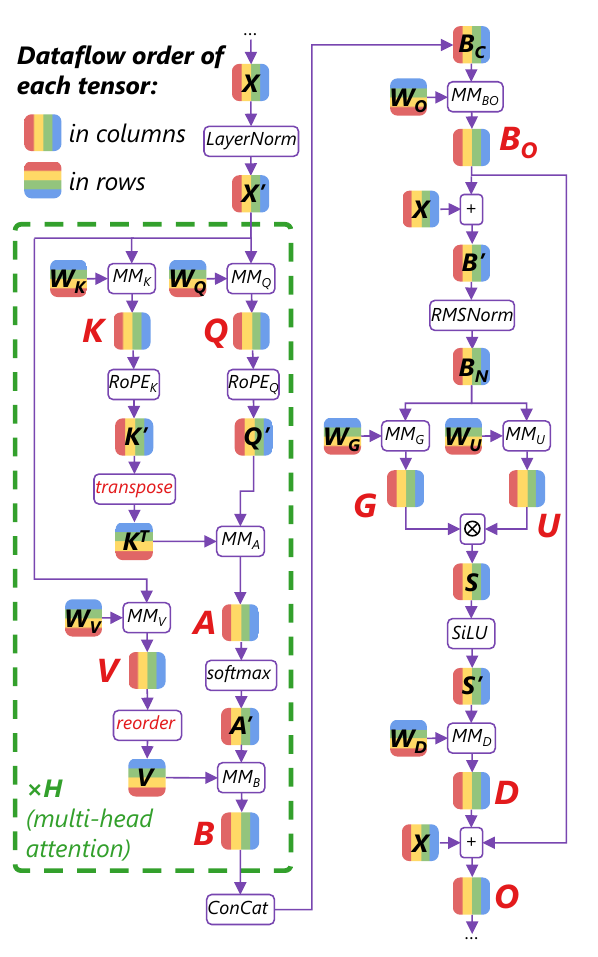}
    \caption{\footnotesize The dataflow graph of a transformer block in {\tt LLaMA}.}
    \label{fig:first_page:graph}
\end{subfigure}
\begin{subfigure}[b]{0.2\textwidth}
\begin{subfigure}[b]{\textwidth}
    \includegraphics[width=\textwidth]{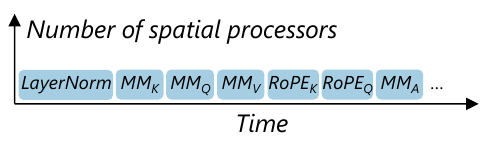}
    \caption{\footnotesize A schedule of the block for a non-dataflow accelerator.}
    \label{fig:motivation:nodf}
\end{subfigure}
\begin{subfigure}[b]{\textwidth}
    \includegraphics[width=\textwidth]{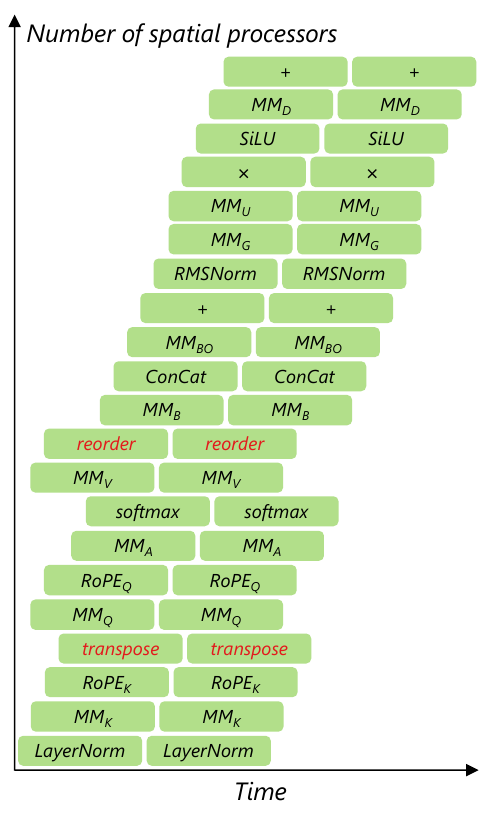}
    \caption{\footnotesize A schedule of the block for a dataflow accelerator.}
    \label{fig:motivation:df}
\end{subfigure}
\end{subfigure}
\caption{
An example of mapping {\tt LLaMA} onto a dataflow accelerator. 
The large variances of each activation cross different layers in (a) motivate us to use quantization with MX formats in (c). We achieve mixed-precision quantization in (b) and map the model onto a dataflow architecture in (d). The dataflow schedule exploits task-level parallelism in (f), leading to a higher throughput compared to a non-dataflow schedule in (e). \textit{The proposed MASE compiler provides a fully automated and efficient approach to exploring software and hardware optimizations for MX formats.}
}
\label{fig:motivation}
\end{figure*}
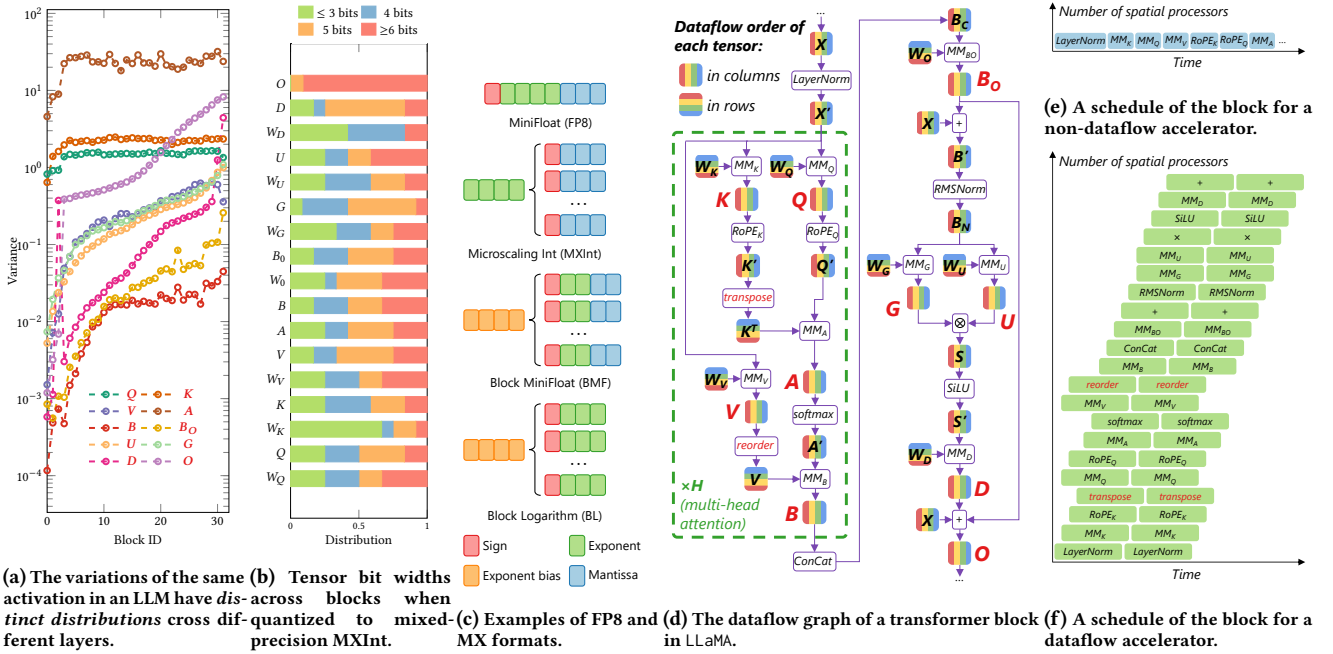

{\bf Background: Large numerical variation in activation values motivates quantization with new data formats for efficient LLM inference.}
Existing LLM hardware accelerator designs typically quantize each tensor into one of two common number representations: fixed-point formats, such as {\tt int8}, and floating-point formats, such as {\tt FP8}. Both representations have their merits: operations with fixed-point formats simplify circuit design, while operations with floating-point formats can enhance accuracy when a wider dynamic range of value is required. LLM quantization is more challenging because it has both a large variation of values and high computational complexity. For example, the variance of all activations in each transformer block of {\tt LLaMA} averaged across all data points in the {\tt Wikitext2} dataset is plotted in Fig.~\ref{fig:motivation:dist}. This plot highlights that the variances change drastically for different tensors in different layers. 
For example, variances increase in deeper layers with significant changes up to 7624$\times$ (the variable {\color{jcred} \bf $D$} in Fig.~\ref{fig:motivation:dist}). Also, variances significantly vary between tensors, even if they are within the same layer; for example, {\color{jcred} \bf $A$} and {\color{jcred} \bf $D$} exhibit a 7902-fold variance difference at layer 0.
This observation motivates us to explore efficient data formats that combine the best of both worlds of fixed-point and floating-point formats.

{\bf Design Opportunities: Microscaling (MX) formats have shown initial promising results in LLM quantization.} 
MX formats are a class of data representations that allow a block of values to share certain components of their data formats, as illustrated in Fig.~\ref{fig:motivation:types}, leading to efficient memory size. Table~\ref{tab:motivation} gives a comparative overview of different MX formats against other arithmetic types when prototyped on FPGAs. Among all data formats, the MXInt format, a subclass of MX formats, offers an advantage in achieving a favorable balance between minimizing accuracy loss and optimizing hardware efficiency. 

{\bf Problem: Existing approaches require manual effort to explore custom data formats for LLM accelerator designs.}
Although MX formats have recently been standardized by AMD, Arm, Intel, Meta, Microsoft, NVIDIA, and Qualcomm~\cite{mxfp}, the exploration of MX formats on hardware accelerators remains limited. A major reason is that there is no tool available to explore these formats for hardware accelerator designs. In practice, significant manual effort is spent on iterations between software model and hardware mapping to determine an optimized co-design. Existing work on MX quantization treats each layer equally and applies the same quantization to all tensors~\cite{darvish2020pushing, rouhani2023shared}. This reduces the design space but also misses opportunities to perform model-specific optimizations for a given LLM, potentially making the optimal hardware design unreachable. 

In order to tackle the problems above, our work aims to solve the following challenges:
\begin{description}[leftmargin=!]
\item[1) Efficiency:] How should one efficiently explore fine-grained quantization of an LLM using custom data formats?
\item[2) Hardware awareness:] How should one determine a quantization solution that leads to an efficient hardware design? 
\end{description}
The efficiency here means minimal design effort, avoiding re-implementing optimizations from scratch for a new data format. Existing optimization algorithms originally for existing data formats may be reused to explore design opportunities for new data formats. These optimizations must take hardware intrinsics into account, leading to a hardware-friendly solution.

{\bf Solution: We propose a novel co-design compiler named MASE to explore custom MX formats for efficient LLM inference on dataflow accelerators.} Specifically, MASE provides an efficient co-design intermediate representation (IR) named {\tt MASE IR} to explore software and hardware co-design with custom data formats. 
A key novelty of {\tt MASE IR} is that it orchestrates existing optimization techniques for traditional data formats to explore hardware optimization opportunities for custom data formats. To efficiently exploit fine-grained mixed-precision custom MX quantization, MASE maps an LLM onto dataflow architectures where each tensor precision can be tailored at the bit level. 

To our knowledge, MASE is the first approach to dataflow hardware design using mixed-precision MX formats. Our main contributions are as follows: 
\begin{itemize}
    \item an end-to-end compiler that automatically determines a mixed-precision MX quantization for a given LLM for mapping onto an efficient dataflow hardware accelerator;
    \item an efficient orchestration method using a co-design IR to explore both software and hardware designs for custom data formats such as MX formats;
    \item an open-source library of parameterized hardware operator designs using MX formats and their evaluation model at the source level for mixed-precision quantization search and efficient hardware generation; and
    \item over a set of LLM families and datasets, our approach attains on average 24\% in $\Delta$ accuracy with an overhead of 3\% in area efficiency compared to designs using 8-bit fixed-point numbers.
\end{itemize}
The rest of the paper is organized as follows. Section~\ref{sec:motivation} provides a motivating example for exploiting custom MX formats in LLM hardware accelerator design; Section~\ref{sec:methodology:mase_ir} describes {\tt MASR IR} and its optimization orchestration; Section~\ref{sec:method:mxint_search} describes our modeling of MXInt formats for efficient quantization search; Section~\ref{sec:experiments} evaluates our design over a set of state-of-the-art LLM models; and Section~\ref{sec:background} reviews related work on block-based quantization, hardware compilers for LLM inference, and hardware accelerator designs.

\section{Motivating Example}
\label{sec:motivation}

In this section, we begin by introducing dataflow hardware architectures and their optimization opportunities for fine-grained quantization. We also provide an overview of three MX formats and evaluate their performance in quantizing LLMs. 

{\bf Why Dataflow Accelerators?}
Dataflow hardware accelerators are specialized hardware architectures that drive operations using the presence of input data, leading to parallelized execution of coarse-grained tasks across several spatial processors. Fig.~\ref{fig:motivation:nodf} illustrates a schedule running on a non-dataflow architecture, such as Von-Neumann architecture, where only one task is executed at a time. This means that all the hardware resources are exploited for each task, leading to a low latency, however, these tasks are sequentially executed in time. On the other hand, a dataflow architecture exploits spatial parallelism among these tasks, leading to a schedule in Fig.~\ref{fig:motivation:df}. Such a parallelism, also known as pipelining, leads to high data throughput, however, the latency is sub-optimal since the hardware resources are shared among different tasks. 

We focus on dataflow architectures because each spatial processor can be tailed for the task that it computes, while non-dataflow architectures require general spatial processors for all tasks. This leads to minimal instruction overhead and design opportunities for fine-grained customization down to the bit level. Targeting a dataflow architecture allows us to simplify the hardware design problem, such as excluding control flow design, and focus on data-specific hardware optimizations.

\begin{figure*}
\centering
\begin{subfigure}[b]{0.45\textwidth}
\centering
\begin{subfigure}[b]{\textwidth}
\centering
\begin{lstlisting}[language=python, escapeinside={(*@}{@*)}, linewidth = \textwidth]
class toy(torch.nn.Module):
    ... 
    # Before quantization, all values are in FP32
    def forward(self, x):
        x0 = flatten(x)
        x1 = relu(linear(x0))
        return x1
\end{lstlisting}
\caption{\footnotesize Input PyTorch model.}
\label{list:pytorch}
\end{subfigure}
\begin{subfigure}[b]{\textwidth}
\centering
\begin{lstlisting}[language=maseir, escapeinside={(*@}{@*)}, linewidth = \textwidth]
toy():
    %x: MXInt((16,2),8,7) = in{}
    %x0: MXInt((16,2),8,7) = FlattenOp(%x) {...}
    %l1: MXInt((16,2),8,7) = LinearOp(%x0)
      [weight: MXInt((16,2),8,3), 
      bias: MXInt((16,2),8,3)] {...}
    %x1: MXInt((16,2),8,7) = ReLuOp(%l1) {...}
    return x1
\end{lstlisting}
\caption{\footnotesize Quantized model in {\tt MASE IR}.}
\label{list:maseir}
\end{subfigure}
\end{subfigure}
\hfill
\begin{subfigure}[b]{0.53\textwidth}
\centering
\includegraphics[width=\textwidth]{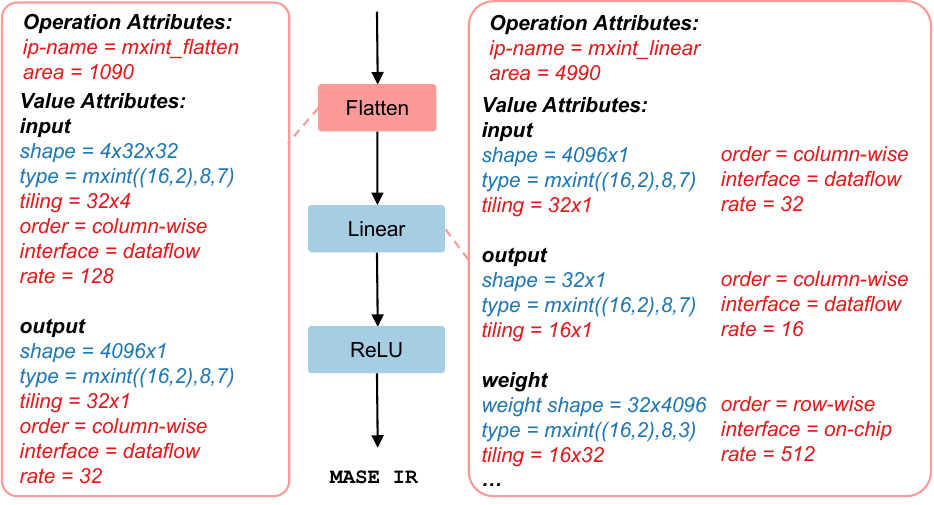}
\caption{\footnotesize A graph view of operation and value attributes of the reorder and linear operation in the model. Here we highlight {\color{jcblue} software-specific} and {\color{jcred} hardware-specific} attributes.}
\label{fig:mase_op_attr}
\end{subfigure}
\caption{A toy model in {\tt MASE IR} after quantization and hardware parallelism.}
\label{fig:code_example}
\end{figure*}

For example, Fig.~\ref{fig:first_page:graph} illustrates a transformer block of {\tt LLaMA} mapped onto a dataflow architecture. In an LLM, tensors are often large, and their computation cannot be fully parallelized for given available hardware resources. Instead, they are partitioned into tiles and streamed into the hardware in a deep pipeline. The streaming orders of these tiles depend on the dataflow hardware operator behaviors. In the figure, tensors are streamed either in a row-by-row or column-by-column order. There are also dataflow-specific operators in the hardware, such as `transpose' and `reorder`, to switch the streaming order in between at run-time. In this work, we take this hardware architecture as a starting point, and explore efficient mixed-precision quantization search at the tensor level using MX formats. For example, Fig.~\ref{fig:motivation:prec} illustrates the bitwidth distribution of the activations and weights in Fig.~\ref{fig:motivation:dist} across transformer blocks using our work.

{\bf What are MX Formats?}
MX formats allow a block of values to share certain components of a data format as a scaling factor~\cite{mxfp}. The scaling factor enables individual values to represent larger dynamic ranges compared to traditional integers. The elements then provide a high-precision representation of values within the range specified by the scaling factor. MX formats could further reduce the average bits per value due to the sharing of the scaling factor in the block. A key requirement is that all the elements in a block must be within the same range specified by the scaling factor. 

By means of examples, we compare three MX formats with standard floating-point formats in Fig.~\ref{fig:motivation:types}. A standard floating-point format contains four components: a sign bit, an exponent, a mantissa, and an exponent bias~\cite{kahan1996ieee}. A common one used for ML is MinFloat (FP8) proposed by Sun \textit{et al.} \cite{sun2019hybrid}, as illustrated at the top of the figure. It has 8 bits overall, and the exponent bias in this format is set as a fixed constant of 7. 
We now introduce three MX formats.
First, the {\em Microscaling Integers (MXInt)} format, also known as the Block floating-point (BFP) format~\cite{kalliojarvi1996roundoff}, shares the exponent in a block. The shared exponent bounds the range of values in the block and works well for values with small typical variation between magnitudes of the components in a block. 
Second, the {\em Block Minifloat (BMF)}~\cite{fox2021block} format shares the exponent bias in a block. This representation achieves high precision and range simultaneously, albeit with a larger quantization error around the medium of its range compared to the standard floating-point format. It is potentially suitable for values in a multi-modal distribution, efficiently representing values close to a peak within a block. 
Finally, the {\em Block Logarithm (BL)} format~\cite{miyashita2016convolutional} strips out the mantissa and shares the exponent bias, resulting in values that are always powers of two. This contrasts with MXInt and is suitable for values with large dynamic ranges.

Table~\ref{tab:motivation} shows the quantization results using different arithmetic formats for {\tt LLaMA} on {\tt Wikitext2}. To ensure fairness, all the arithmetic types have an average bitwidth of 8 bits. We evaluate them using three metrics, perplexity, memory density, and arithmetic density. The memory and arithmetic densities represent the normalized average values per bit and normalized average area per arithmetic operation compared to {\tt FP32}~\cite{darvish2020pushing}. Both memory and arithmetic densities are derived from our post-routing hardware GEMM implementation using these arithmetic types. From the table, we made following observations. First, traditional 8-bit fixed-point ({\tt Int8}) quantization achieves decent memory and arithmetic density but suffers from a significant increase in perplexity. Second, FP8 achieves the best hardware efficiency, with an increase in perplexity. Finally, MX formats, such as MXInt, have competitive memory density and arithmetic density, and can preserve low perplexity. This motivates us to explore custom MX formats for LLM quantization and further improve hardware area efficiency with minimal precision loss. 

{\bf What is the most efficient data format and its precision for quantizing an LLM?} 
Given an LLM, the proposed compiler, MASE, automatically finds a mixed-precision MX quantization solution and maps it into an efficient dataflow accelerator for inference. 
In the rest of the paper, we show how to exploit our proposed abstraction {\tt MASE IR} for efficient design exploration using MX formats.



\section{MASE Intermediate Representation}
\label{sec:methodology:mase_ir}

Existing hardware compilers for ML accelerators suffer from two main problems. First, they focus on operations with fixed-point and floating-point formats. Designers need to manually re-implement the whole design from scratch when mapping models with custom data formats. Second, the hardware-aware IR in those hardware compilers, such as LLVM~\cite{lattner2004llvm} and MLIR~\cite{lattner2021mlir}, do not preserve backforward propagation functions. This means that a software model lowered into such an IR can no longer be further trained. When training is required, designers need to restart from a state in the software flow and may abandon all applied hardware optimizations if an additional software optimization is applied. In order to overcome these two challenges, we propose {\tt MASE IR}, a hardware-aware and `trainable' software intermediate representation that describes both a software model and the corresponding hardware accelerator architecture. {\tt MASE IR} provides an efficient interface for users to integrate custom data formats for hardware exploration, and also keeps backward propagation functions so that the model can be trained or fine-tuned in hardware optimization cycles.

\begin{table}[]
    \centering
    \caption{Key MASE passes used in this work. MASE contains 44 analysis and optimization passes. All these passes target different granularities varying from the model level to the bit level. These passes are general and type independent, which opens up opportunities for optimizing new data formats. Here we highlight \sw{software-specific} and \hw{hardware-specific} components.}
    \label{tab:mase_ir_attr}
{\footnotesize
\begin{tabular}{l p{6cm}}
\toprule
Names & Descriptions \\
\midrule
\sw{profile} & Profile variation of values for a given dataset, used to define the quantization search space. \\
\midrule
\sw{quantize} & Quantize a given model based on an input configuration, used to perform tensor-level mixed-precision quantization for a given data format. \\
\midrule
\hw{parallelize} & Exploit resource-constrained hardware parallelism based on a given hardware target, leading to a hardware design with high area efficiency. \\
\midrule
\texttt{evaluate} & Evaluate the hardware design based on a given expression of cost function, taking both model accuracy and area efficiency as arguments. \\
\midrule
\texttt{search} & Orchestrate existing search algorithms, such as random search and Tree-structured Parzen Estimator (TPE), to explore quantization search. \\
\midrule
\hw{emit} & Translate a co-design in {\tt MASE IR} into a dataflow hardware accelerator in SystemVerilog. \\
\bottomrule
\end{tabular}}
\end{table}
\begin{figure*}
    \centering
    \includegraphics[width=\textwidth]{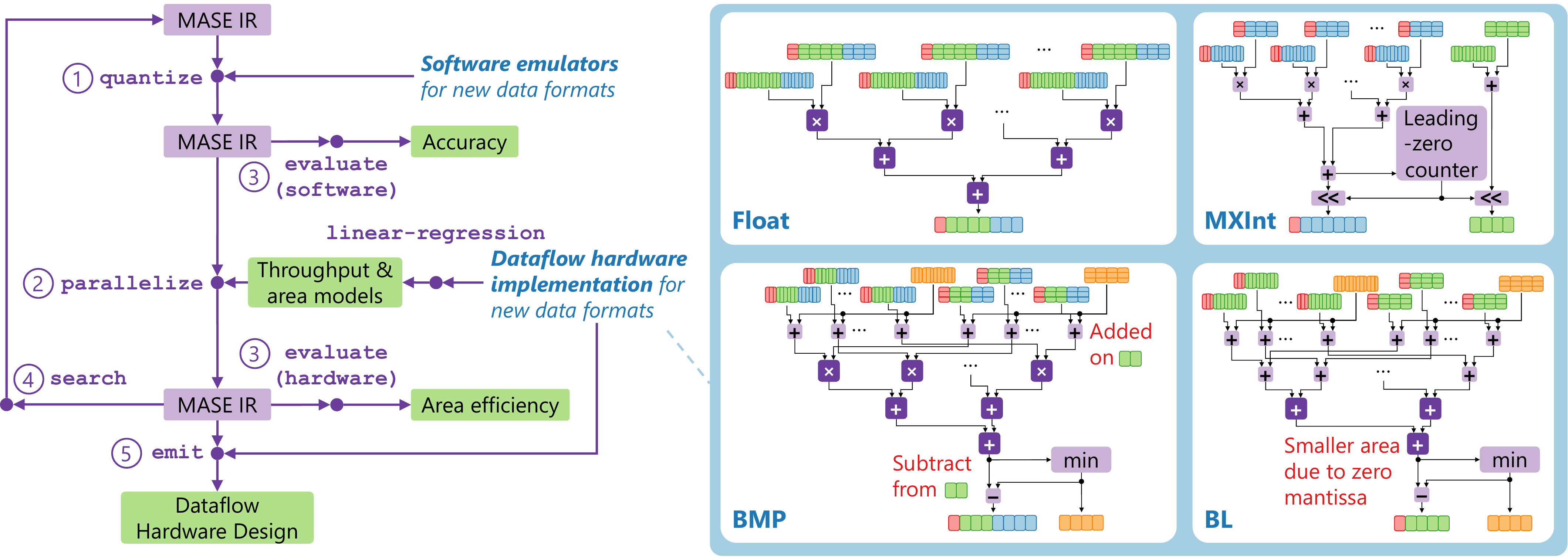}
    \caption{Orchestration of existing tools for new data formats exploration. Given both software and hardware specifications for a new data format, MASE automatically explores resource-constrained quantization search for a given LLM.
    }
    \label{fig:orchestration}
\end{figure*}

Like most IR languages, the syntax of {\tt MASE IR} follows the traditional static single-assignment form~\cite{cytron1991efficiently}. The SSA form already provides a dataflow-like representation of a model. This enables direct translation into a dataflow hardware representation where each software module is mapped into a hardware component as illustrated in Fig.~\ref{fig:first_page:graph} and connected to other components using handshake interface. An operation in {\tt MASE IR} contains a set of components, arguments, results, parameters and attributes:

\texttt{\footnotesize result: type = operator(arg: type, ...) [param: type, ...] \{attr, ...\}}

\noindent
{\tt MASE IR} is general for representing any ML model to explore optimizations with custom data formats. For example, Fig.~\ref{list:pytorch} illustrates a toy model that contains a {\tt Linear} operation followed a {\tt ReLU} function. The input tensor is flattened before being sent to the linear operation. Fig.~\ref{list:maseir} represents the quantized toy model instance in {\tt MASE IR}. All the tensors in the model including both activations and parameters are quantized in custom-precision MXInt formats for smaller bitwidths. For example, a type of MXint((16, 2), 8, 7) means that the elements of the tensor shares an 8-bit exponent for every block of size 16 by 2, and every element has a 7-bit mantissa. A model in {\tt MASE IR} also carries detailed hardware design attributes for parallelism exploration, as illustrated in Fig.~\ref{fig:mase_op_attr}. The operation attributes specify which hardware IP block is used for exploration and its estimated circuit area. The value attributes describe each dataflow edge in Fig.~\ref{fig:first_page:graph}. As illustrated in Fig.~\ref{fig:mase_op_attr}, these include the shape of streaming tiles, the streaming order, hardware data interface, and estimated throughput. This allows the model optimizer to interface existing tools to exploit hardware parallelism. 

\subsection{Passes for Quantization Search and Dataflow Optimization}

MASE contains a large set of passes targeting analysis and optimizations at different granularities ranging from the model level to the bit level. To minimize additional development for the new format, all MASE passes are type-independent, so that they can be orchestrated for optimizations of any data format. A set of key passes used in this work are listed in Table~\ref{tab:mase_ir_attr}. The left of Fig.~\ref{fig:orchestration} provides an example of quantization search flow in {\tt MASE IR}. 
In this example, the toy model in Fig.~\ref{list:pytorch} is translated into {\tt MASE IR} from PyTorch. The MASE front-end automatically performs model analysis and initializes software attributes when constructing {\tt MASE IR}, such as tensor shapes and initial data types. For complex models, implicit dataflow-specific operations may also be inserted, such as `reorder' in Fig.~\ref{fig:first_page:graph}. For simplicity, here we focus only on the toy model.
{\textcolor{jcpurple}{\large \textcircled{\raisebox{-0.9pt}{\normalsize 1}}}}
The model is quantized by the quantize pass using a set of user-defined precisions, which supports both post-training quantization (PTQ) and quantization-aware training (QAT). For this example, the model is quantized into MXInt format at the tensor level. The parameters have a lower bitwidth compared to activations because they are less sensitive. 
{\textcolor{jcpurple}{\large \textcircled{\raisebox{-0.9pt}{\normalsize 2}}}}
Then the parallelize pass exploits hardware parallelism for the quantized model. Given a hardware resource budget, it automatically explores the most efficient stream tile sizes for each layer, leading to an optimized overall throughput. The hardware mapping solution contains several hardware design parameters illustrated in Fig.~\ref{fig:mase_op_attr}. 
{\textcolor{jcpurple}{\large \textcircled{\raisebox{-0.9pt}{\normalsize 3}}}}
Since {\tt MASE IR} contains both software and hardware design parameters, the accuracy of the model and the area efficiency of the final hardware can be estimated by program analysis at the source level using the evaluate pass. These design constraints form a hardware-aware cost function, which could guide exploration of both following software and hardware optimizations.
{\textcolor{jcpurple}{\large \textcircled{\raisebox{-0.9pt}{\normalsize 4}}}}
Guided by the cost function, the search pass iterates quantization and hardware parallelism to find an efficient mixed-precision quantization for a given model. It orchestrates existing search algorithms, such as TPE (Tree-structured Parzen Estimator) \cite{ozaki2020multiobjective}, for efficient exploration with custom data formats. 
{\textcolor{jcpurple}{\large \textcircled{\raisebox{-0.9pt}{\normalsize 5}}}}
After a given number of iterations, the model is then mapped into a dataflow hardware design in SystemVerilog. The emit pass performs direct translation without any program analysis because all the hardware design parameters of the model are accessible in {\tt MASE IR}.

\subsection{Pass Orchestration for Custom Data Formats}

MASE is general and allows for seamless integration of new data formats for resource-constrained quantization. The right of Fig.~\ref{fig:orchestration} shows our optimization orchestration flow, where users only need to add the blue part of the code for a new data format. There are two components to be added for a new data format, software emulators and hardware components. First, software emulators specify how to quantize and dequantize the value between the given format and floating-point numbers. For each operation, the input data is first quantized into a given input precision. Then the same existing operation in PyTorch is orchestrated to carry out the calculation in floating-point numbers. The results in floating-point numbers are further quantized to a given output precision. A main advantage of such an approach is that MASE can orchestrate existing floating-point operations to emulate operations with custom data formats without re-defining all the operations from scratch. The software emulators provide a fast evaluation of model performance in accuracy, which guides further software optimizations, such as the iterative quantization search. 

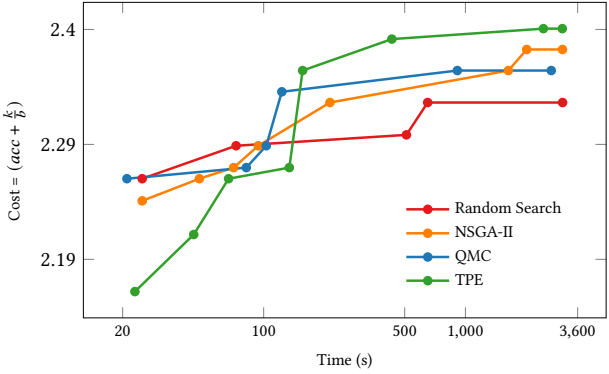
\begin{figure}
    \centering
    \begin{tikzpicture}[thick,scale=0.95, every node/.style={scale=0.95}]
\pgfplotsset{every x tick label/.append style={font=\footnotesize}, compat=1.3}
\begin{semilogyaxis}[
    height=60mm,
    width=0.5\textwidth,
    xlabel={\footnotesize Time (s)},
    ylabel={\footnotesize Cost = $(acc + \frac{k}{b})$},
    xmin=0, 
    xtick={20, 100, 500, 1000, 3600},
    ymode=log, log ticks with fixed point, xmode=log,
    legend cell align={left},
    legend style={at={(0.6,0.4)},anchor=north west, draw=none, fill=none, font=\footnotesize},
    ]

\addplot[jcred, draw=jcred, mark=*, line width=1pt, mark options={scale=0.7}] table
[x=t_rs, y=rs, col sep=space] {alg_rs.dat};
\addlegendentry{Random Search}

\addplot[jcorange, draw=jcorange, mark=*, line width=1pt, mark options={scale=0.7}] table
[x=t_nsga, y=nsga, col sep=space] {alg_nsga.dat};
\addlegendentry{NSGA-II}

\addplot[jcblue, draw=jcblue, mark=*, line width=1pt, mark options={scale=0.7}] table
[x=t_qmc, y=qmc, col sep=space] {alg_qmc.dat};
\addlegendentry{QMC}

\addplot[jcgreen, draw=jcgreen, mark=*, line width=1pt, mark options={scale=0.7}] table
[x=t_tpe, y=tpe, col sep=space] {alg_tpe.dat};
\addlegendentry{TPE}

\end{semilogyaxis}
\end{tikzpicture}    
    \caption{Evaluation of search algorithms for {\tt OPT125M} on {\tt sst2}. MASE orchestrates existing search algorithms to explore resource-constrained quantization with mixed-precision MXInt formats. The cost function is shown on the y label (described in Section~\ref{sec:method:mxint_search}. $acc$ = accuracy, $b$ = average bitwidth. $k$ is a hyperparameter to normalize costs. We observed that TPE is the most efficient search algorithm for MXInt quantization.}
    \label{fig:search_algorithms}
\end{figure}

Second, designs for such hardware operators can be diverse, making it challenging for existing tools to estimate hardware results. To restrict the design space, the hardware designs of operators with new data formats are required for architecture exploration and evaluation. Each component is provided as a Verilog template of a dataflow component with a set of parameters for data parallelism, such as input stream tile sizes. For example, the right of Fig.~\ref{fig:orchestration} illustrates a high-level view of a dot product operation with four data formats illustrated in Fig.~\ref{fig:motivation:types}. The light purple blocks represent fixed-point or logic operators which have small area, and the dark purple blocks represent floating-point operators which have large area. Compared to the traditional floating-point operator on the top left, the {\em MXInt} operator saves significant area by reusing the results of the shared exponent in the block because one of the main area costs of a floating-point operator is the dynamic shift hardware unit~\cite{coward2023automating}; the {\em BMF} operator, on the other hand, requires more circuit area to calculate values with the shared exponent bias, while expanding the dynamic range of each element; the {\em BL} operator saves area from the BMF operator by stripping out operators for the mantissas, leading to a low precision in a small range. 

With the provided hardware templates, MASE automatically explores hardware designs by sweeping the parallelism parameters. This is a one-off process, and a regression model enables MASE to estimate the overall throughput and total circuit area at the source level.
MASE orchestrates existing optimization algorithms for dataflow architecture exploration and uses the regression model to guide the process. This leads to an efficient co-design with both high area efficiency and accuracy.
The provided Verilog templates are also used for final hardware generation. Because all hardware components are implemented in a dataflow architecture, they can be directly integrated into the design by connecting the handshake interface. 
{\em The MX software emulators and hardware components will be open source as well as the MASE tool.}

\begin{table}
    \centering
\caption{Line-of-Code comparison for {\tt OPT}. {\tt MASE IR} provides an efficient representation of an ML model at the module level and enables fast compilation time compared to existing hardware design IR, such as MLIR {\tt affine}. DAG = code in directed acyclic graph.}
\label{tab:loc}
\resizebox{0.48\textwidth}{!}{%
\begin{tabular}{lrrrrr}
\toprule
\multicolumn{1}{c}{Models} & \multicolumn{1}{l}{\makecell[c]{MLIR {\tt affine}\\DAG Size}} & \multicolumn{1}{l}{\makecell[c]{Codegen\\Time}} & \multicolumn{1}{l}{\makecell[c]{MASE IR\\DAG size}} & \multicolumn{1}{l}{\makecell[c]{Codegen\\Time}} & \makecell[c]{Code size\\$\times$} \\
\midrule
{\tt OPT-125M} & 1.9M & 1 week  &  61  & 23s  &  31.1k\\
{\tt OPT-350M} & 1.7M & 2 weeks &  86  & 63s  & 19.7k\\
{\tt OPT-1.3B} & 1.7M & $>$4 weeks &  86  & 112s & 19.7k\\
{\tt OPT-2.7B} & 1.9M & $>$4 weeks &  101  & 217s & 18.8k\\
{\tt OPT-6.7B} & 2.3M & $>$4 weeks &  101  & 467s & 22.8k\\
\bottomrule
\end{tabular}
}
\end{table}

\subsection{Scalability Analysis}

{\bf Optimization orchestration enables users to explore a range of new data formats at scale.}
This minimizes development time and effort and allows users to focus on exploring efficient data formats using advanced algorithms. For example, Fig.~\ref{fig:search_algorithms} evaluates four well-known search algorithms orchestrated by MASE for resource-constrained mixed-precision MXInt quantization on {\tt OPT-125M}, Random Search, Non-dominated Sorting Genetic Algorithm II (NSGA-II)~\cite{deb2002fast}, Quasi-Monte Carlo (QMC) sequences~\cite{bergstra2012random}, and Tree-structured Parzen Estimator (TPE)~\cite{bergstra2011algorithms}. Random Search is an elementary method that involves exploring the solution space by generating random configurations. The NSGA-II method is a multi-objective optimization algorithm that operates on the principles of evolutionary algorithms, wherein a population of candidate solutions undergoes a process of evolution through selection, crossover, and mutation operators. The QMC method is a class of numerical integration techniques used for high-dimensional problems where the use of traditional Monte Carlo (MC) methods would be computationally infeasible. The TPE method is a Bayesian optimization algorithm that models the dependency between hyperparameters to efficiently discover the promising areas of the hyperparameter search space. 

In comparison, random search serves as a straightforward baseline but has the minimum change between the starting design point and the final design point. This is due to the lack of a guided search strategy. NSGA-II has a slightly larger change and leads to a better design, efficiently trading off between the accuracy and memory size. The QMC method has the fastest search speed but results in a sub-optimal design. TPE, although it has the worst starting point, can be effectively improved over time and results in the best design among all the algorithms. The average search times for all these algorithms are close. This suggests that the TPE is the most efficient algorithm to search for mixed-precision MXInt quantization, so we use the TPE algorithm in our experiments. 

{\bf {\tt MASE IR} also provides a compact representation for exploration of large models.}
It efficiently expresses the dataflow architecture for a large model up to billions of parameters and achieves significant scalability improvements in compilation time compared to existing hardware IRs. 
Here we compare {\tt MASE IR} with the MILR {\tt affine} dialect~\cite{lattner2021mlir}, a commonly used hardware compilers~\cite{ye2022scalehls, zhao2022polsca, cheng2023seer}. Table~\ref{tab:loc} compares the code size of MLIR and {\tt MASE IR} in directed acyclic graph (DAG) size. In the table, we observe that MASE IR has shown a significantly smaller code size than MLIR {\tt affine}, because it expresses operations at the module level, hiding the instruction-level details from users. The detailed MLIR enables a finer-grained hardware optimization but the code size overhead causes the compilation time to increase exponentially. In MASE, the optimizations focus on the modules, where instruction-level optimizations are manually carried out in the Verilog templates and linked to the tunable parameters. This reduces the compilation complexity while preserving high hardware efficiency. In addition, the hardware attributes in {\tt MASE IR} also enables MASE to interface with these hardware compilers to explore instruction-level optimizations in MLIR by loading their generated hardware designs instead of Verilog templates for regression modeling in Fig.~\ref{fig:orchestration}.

\section{Mixed-Precision MXInt Quantization}
\label{sec:method:mxint_search}

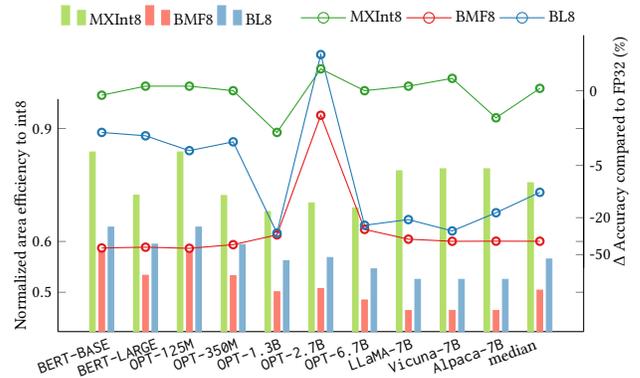
\begin{figure}[t]
\centering
\pgfplotsset{compat=1.3}
\begin{tikzpicture}[thick,scale=0.7, every node/.style={scale=0.7}]
\pgfplotsset{set layers}
\begin{axis}[
    width=0.65\textwidth,
    ybar, height=6cm, 
    ymode=log,
    legend style={at={(-0.05,1.5)},anchor=west, draw=none, fill=none},
    legend columns = 3,
    ymin=0, 
    ymax=1,
    ylabel={ Normalized area efficiency to int8},
    ytick={0.5, 0.6, 0.9},
    axis x line*=bottom,
    bar width=3,
    legend cell align={left},
    log ticks with fixed point,
    major x tick style = transparent,
    symbolic x coords={
bertbase,bertlarge,opt125m,opt350m,opt13b,opt27b,opt67b,llama7b,vicuna,alpaca,mean
},
    xticklabels={{\tt BERT-BASE},{\tt BERT-LARGE} ,{\tt OPT-125M} ,{\tt OPT-350M} ,{\tt OPT-1.3B} ,{\tt OPT-2.7B} ,{\tt OPT-6.7B}, {\tt LLaMA-7B}, {\tt Vicuna-7B}, {\tt Alpaca-7B}, {median}},
    xtick=data,
    xticklabel style={xshift=-15pt, yshift=5pt, rotate=20},
    nodes near coords,
    point meta=explicit symbolic,
    nodes near coords style={
            align=center,
        },
    ]

\addplot[jcgreenl, draw=jcgreenl, fill=jcgreenl] table
[x=models, y=mxint8, col sep=space] {table1.dat};
\addlegendentry{ MXInt8}

\addplot[jcredl, draw=jcredl, fill=jcredl] table  
[x=models, y=bmf8, col sep=space] {table1.dat};
\addlegendentry{ BMF8}

\addplot[jcbluel, draw=jcbluel, fill=jcbluel] table
[x=models, y=bl8, col sep=space] {table1.dat};
\addlegendentry{ BL8}

\end{axis}

\begin{axis}[
    width=0.65\textwidth,
    height=6cm, 
    axis y line*=right,
    legend style={at={(0.6,1.5)},anchor=west, draw=none, fill=none},
    legend columns = 3,
    ylabel={$\Delta$ Accuracy compared to FP32 (\%)},
    ytick={-1.707570176, -1.322219295, -0.7781512504, 0, 0.7781512504},
    yticklabels={-50, -20, -5, 0, 5, 20, 50},
    legend cell align={left},
    log ticks with fixed point,
    xticklabels=\empty,
    nodes near coords,
    point meta=explicit symbolic,
    nodes near coords style={
            align=center,
        },
    hide x axis,
    ]

\addplot[jcgreen, draw=jcgreen, mark=o] table
[x=model, y=mxint8, col sep=space] {table1a.dat};
\addlegendentry{ MXInt8}

\addplot[jcred, draw=jcred, mark=o] table  
[x=model, y=bmf8, col sep=space] {table1a.dat};
\addlegendentry{ BMF8}

\addplot[jcblue, draw=jcblue, mark=o] table
[x=model, y=bl8, col sep=space] {table1a.dat};
\addlegendentry{ BL8}

\end{axis}

\end{tikzpicture}
\caption{Evaluation of three MX data formats for quantizing LLMs on {\tt sst2}. The area efficiency results are plotted relative to int8 results (higher means better). The accuracy are represented as its difference with the accuracy using FP32 (higher means better). To ensure fairness, all the formats have a block size of 32 that contains an 8-bit shared component and 8-bit local components, leading to an average bitwidth of 8 bits. Overall, MXInt has shown both high area efficiency and high accuracy for LLM quantization. }
\label{fig:compare_mx_types}
\end{figure}

We will now demonstrate the exploration of MX formats quantization using MASE. Insights derived from Table~\ref{tab:motivation} indicate that among the evaluated MX formats, MXInt is better suited for quantizing the {\tt LLaMA} model.
Fig.~\ref{fig:compare_mx_types} further verifies the same phenomenon across $10$ different LLMs. To ensure fairness, every model is quantized to a given format with an average bitwidth of 8 bits, and mapped onto dataflow hardware using the same hardware optimizations. The bars represent the normalized hardware area efficiency of each design to the standard dataflow hardware implementation using 8-bit fixed point numbers (int8). Larger area efficiency means more efficient hardware design. The curves plot the difference in accuracy compared to the accuracy of the model in FP32. A larger accuracy difference means higher accuracy. 

{\bf MXInt is the most amenable for quantizing LLMs.}
In general, MX formats require more complex circuit design, leading to lower area efficiency compared to int8; however, their dynamic ranges lead to better accuracy. 
Among these MX formats, we observed that the MXInt format achieves both the best area efficiency and the best accuracy for most models.\footnote{There is a special case for {\tt OPT-2.7B} because the original model in FP32 already has a low zero-shot accuracy.} This suggests that for most values in these LLMs, their elements in each tenor have small differences from their neighbor elements. We conclude that MXInt is more suitable for quantizing LLMs. In this section, we take MXInt for example, and illustrated our proposed mixed-precision MXInt quantization for LLMs.

{\bf No mixed-arithmetic but mixed-precision quantization.}
Another possibility is to mix arithmetic types as well as mixing precisions, however, the area overhead is large. These formats, such as MXInt and BL, have significant differences in both shared and local elements. The casting function between these formats requires complex dynamic shift operations to align their ranges before any further calculations. This would lead to significant circuit area compared to computing using a single arithmetic type. On the other hand, casting values in different precisions of the same format is more affordable in hardware. Here we take MXInt, for example. Casting mantissas only requires bit extension or truncation, similar to fixed-point numbers. The exponents may require dynamic shift, but the shift operation can be fully unrolled into logic wires at low cost because the bitwidth of mantissas is small. Therefore, mixed-precision MXInt quantization has a low area overhead in type casting between values.

\subsection{Software Design Parameters}
\label{sec:method:software_design_parameters}

The quantization search for MXInt formats involves two sets of constraints, software and hardware. Here we show how to formalize the design parameters and restrict the search space for better search efficiency. Search efficiency entails finding a precise quantization solution with high accuracy and small circuit area using minimal number of trails.

For each parameter or activation value, its MXInt format is a 3-tuple, $(B, e, m)$.
\begin{itemize}
    \item $B \in \mathbb{N}^N$ denotes the shape of the block, indicating how the exponent is shared among elements inside the block. A block usually has two dimensions, where $N = 2$;
    \item $e \in \mathbb{N}$ is the bitwidth of the shared exponent of each block; and
    \item $m \in \mathbb{N}$ is the bitwidth of the mantissa for each element.
\end{itemize}
The average bitwidth $p$ of a value is the sum of the exponent bits per element, the mantissa bits and the sign bit:
\begin{align}
    p = \frac{e}{\Pi B} + m + 1
\end{align}
For example, a value of MXint((16, 2), 8, 7) shown in Fig.~\ref{list:maseir} has an average bitwidth of 8.25. Assume that all blocks have two dimensions. We have the total search space $S$ as follows:
\begin{align}
    S = \mathbb{N}^{4v}
\end{align}
$v$ is the total number of values in the model to be quantized. Each value has four parameters to explore for their precision. Such a search space is significantly huge in a large model that contains hundreds of values. We now show how to efficiently reduce the search space, thus improving search efficiency. 

{\bf Use a unified block shape for all values to reduce the search space.}
A large block size allows an exponent to be shared by more elements, reducing the average bitwidth, and a small block extends the flexibility of element ranges due to relatively more localized scaling factors. Mixed block shapes may achieve finer quantization granularity for each value. However, casting between MXInt formats with different block shapes also requires complex circuits to denormalize and renormalize elements in different blocks. Using a unified block shape is beneficial for reducing quantization search and hardware design complexity. Prior work has shown that a block size of 32 for MXInt can achieve accuracy comparable to {\tt FP16}~\cite{darvish2020pushing, mxfp}. Here we use a block shape of 16 by 2 for all MXInt formats.

{\bf Use a fixed bitwidth for all shared exponents.}
A small exponent leads to a small average bitwidth. However, its effect is negligible when the block size is large. For example, reducing by a bit only causes a reduction of $\frac{1}{32}$ in the average bitwidth for MXInt formats with a block size of 32. Searching for small exponent bits exponentially expands the search space with nearly no benefit. Instead, we use an 8-bit exponent for all MXInt formats.
The design parameters now only have variable mantissa bits, leading to a reduced search space $S'$:
\begin{align}
    S' = \mathbb{N}^{v}
\end{align}
The mantissa bitwidths are essential for quantizing LLMs because the value differences are small. 

{\bf MXInt formats have a smaller search space than fixed-point numbers.}
Our formalization and analysis of MXInt quantization significantly reduce the search space. Compared to mixed-precision fixed-point search, which searches for both total bitwidth and fraction bitwidth for each value (search space of $\mathbb{N}^{2v}$), MXInt quantization searches for mantissa bitwidth of each element, leading to a smaller search space. 

\input{imgs/model_dataset_accuracy}

\subsection{Hardware Design Parameters}
\label{sec:method:hardware_design_parameters}

Hardware optimization techniques for dataflow accelerators have been widely studied in existing hardware compilers~\cite{ye2022scalehls, fahim2021hls4ml, venieris2016fpgaconvnet, umuroglu2017finn}. 
Dataflow hardware optimization techniques typically involve two levels of parallelism. 

{\bf Data Parallelism and Pipelining.}
The streaming tile of each value needs to be efficiently sized. A larger tile exploits more opportunities for spatial parallelism in hardware operators, which also take more hardware resources. The total hardware resources must be efficiently shared among these hardware operators to achieve high overall throughput. For example, different operators may compute at different throughputs because of different hardware behaviors. This means that a set of tile sizes need to be determined for balanced throughput between operators. 

{\bf Memory Allocation.}
Fisrt, most parameter sizes of an LLM are large, taking large memory sizes. These data need to be efficiently allocated either on fast on-chip memory or large off-chip memory. An efficient memory allocation solution must be determined to maximize the efficient utilization of hardware resources and high throughput. Second, the data dependency between operators may cause pipeline stalls, affecting overall throughput. Buffers should be inserted between operators to resolve pipeline stalls to improve throughput.

These design considerations have been widely studied in dataflow hardware architectures for ML inference. Related works~\cite{rucker2024revet, zhao2023sigma, zhang2018dnnbuilder, ye2023hida} propose efficient algorithms to automatically determine an efficient dataflow hardware design, and these can be orchestrated by MASE for efficient hardware exploration. In this work, our scope focuses on optimization orchestration instead of algorithm efficiency. We integrate these hardware design considerations into the same TPE algorithm that runs the quantization search for resource-constrained quantization. The circuit area is obtained from the regression model of hardware operators, and the overall throughput is the minimum throughput among all hardware operators estimated from the regression model.

\subsection{Resource-Constrained Mixed-Precision Search}
\label{sec:method:quantization_search}

Adding hardware design parameters into the quantization search enables efficient software and hardware co-design. The search objective is as follows. 
\begin{align}
    objective : \max (acc + \frac{k}{b} + k'\theta + \frac{k''}{A}) \label{eqn:new_cost}
\end{align}
$acc$ is the model accuracy, $b$ is the average bitwidth of the model, $\theta$ is estimated overall throughput, and $A$ is the estimated total circuit area. $k$, $k'$, and $k''$ are hyperparameters that normalize these design constraints. 

\section{Experiments}
\label{sec:experiments}

We evaluated MASE on ten well-known LLMs from three families, including {\tt BERT}~\cite{devlin2019bert}, {\tt OPT}~\cite{zhang2022opt} and {\tt LLaMA} (including {\tt Vicuna} and {\tt Alpaca})~\cite{touvron2023llama, chiang2023vicuna, miao2023specinfer}. All of them are obtained directly from HuggingFace~\cite{huggingface}. We evaluated the accuracy after mixed-precision {\tt MXInt} quantization on six downstream tasks, including 
{\tt boolq}~\cite{clark2019boolq},
{\tt mnli}~\cite{williams2017broad},
{\tt qnli}~\cite{rajpurkar2016squad}, 
{\tt qqp}~\cite{wang2019glue},
{\tt rte}~\cite{dagan2005pascal},
and {\tt sst2}~\cite{socher2013recursive}.
We evaluate the model accuracy following the same approach proposed by Zhang \textit{et al.}~\cite{zhang2022opt} and Brown \textit{et al.}~\cite{brown2020language}. We use Alveo U250 FPGAs as the target platform for the evaluation of dataflow hardware design, and the version of the Xilinx software used is 2023.1. The throughput results are obtained from on-board measurements. The area and power results were obtained from the Post Place \& Route report in Vivado.

In this section, we compare the accuracy and area efficiency of our mixed-precision MXInt {\bf (MP MXInt)} approach with other metods.
Second, we evaluate the effectiveness of our quantization on accuracy and average bitwidths for different sizes of {\tt OPT} on six downstream tasks. 
Finally, we compare our mixed-precision approach with uniform-precision MXInt, and provide insights for ASIC accelerator designs. 

\begin{figure}[t]
\centering
\pgfplotsset{compat=1.3}
\begin{tikzpicture}[thick,scale=0.7, every node/.style={scale=0.7}]
\pgfplotsset{set layers}
\begin{axis}[
    width=0.65\textwidth,
    ybar, height=6cm, 
    ymode=log,
    legend style={at={(-0.05,1.65)},anchor=west, draw=none, fill=none, cells={align=left}},
    legend columns = 1,
    ymin=0, ymax=2.5,
    ylabel={ Normalized area efficiency to int8},
    ytick={1, 1.3, 1.5, 1.7, 2, 2.4},
    bar width=3,
    legend cell align={left},
    log ticks with fixed point,
    axis x line*=bottom,
    major x tick style = transparent,
    symbolic x coords={
bertbase,bertlarge,opt125m,opt350m,opt13b,opt27b,opt67b,llama7b,vicuna,alpaca,mean
},
    xticklabels={{\tt BERT-BASE},{\tt BERT-LARGE} ,{\tt OPT-125M} ,{\tt OPT-350M} ,{\tt OPT-1.3B} ,{\tt OPT-2.7B} ,{\tt OPT-6.7B}, {\tt LLaMA-7B}, {\tt Vicuna-7B}, {\tt Alpaca-7B}, {median}},
    xtick=data,
    xticklabel style={xshift=-15pt, yshift=5pt, rotate=20},
    nodes near coords,
    point meta=explicit symbolic,
    nodes near coords style={
            align=center,
        },
    ]

\addplot[jcblue, draw=jcblue, fill=jcblue] table
[x=models, y=mixedint, col sep=space] {table1.dat};
\addlegendentry{MP int}

\addplot[jcgreenl, draw=jcgreenl, fill=jcgreenl] table
[x=models, y=mixedint_nodse, col sep=space] {table1.dat};
\addlegendentry{MP MXint (SW-only)}

\addplot[jcgreen, draw=jcgreen, fill=jcgreen] table
[x=models, y=mixedintdse, col sep=space] {table1.dat};
\addlegendentry{MP MXint}

\end{axis}

\begin{axis}[
    width=0.65\textwidth,
    height=6cm, 
    axis y line*=right,
    legend style={at={(0.8,1.6)},anchor=west, draw=none, fill=none},
    legend columns = 1,
    ylabel={$\Delta$ Accuracy compared to FP32 (\%)},
    ytick={-1.707570176, -1.322219295, -0.7781512504, 0, 0.7781512504},
    yticklabels={-50, -20, -5, 0, 5},
    legend cell align={left},
    log ticks with fixed point,
    xticklabels=\empty,
    nodes near coords,
    point meta=explicit symbolic,
    nodes near coords style={
            align=center,
        },
    hide x axis,
    ]

\addplot[jcblue, draw=jcblue, mark=o] table
[x=model, y=mixedint, col sep=space] {table1a.dat};
\addplot[jcgreenl, draw=jcgreenl, mark=o] table  
[x=model, y=mixedint_nodse, col sep=space] {table1a.dat};
\addplot[jcgreen, draw=jcgreen, mark=o] table
[x=model, y=mixedintdse, col sep=space] {table1a.dat};

\addlegendentry{MP int}
\addlegendentry{MP MXint (SW-only)}
\addlegendentry{MP MXint}

\end{axis}

\end{tikzpicture}
\caption{Evaluation of MXInt data formats for quantizing LLMs on {\tt sst2}. The area efficiency results are plotted relative to int8 (higher means better). The accuracy are represented as its difference with the accuracy using FP32 (higher means better). 
}
\label{fig:compare_mx}
\end{figure}
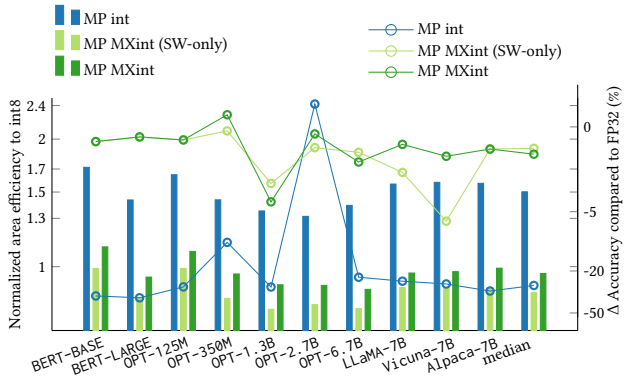

\subsection{Comparison with other quantization approaches}

Here we compared the quality of our co-design with a few baselines. Fig.~\ref{fig:compare_mx} shows the area efficiency of the dataflow hardware designs and model accuracy using different approach. {\bf int8} means quantization using 8-bit fixed-point numbers. {\bf MP int} means mixed-precision quantization using fixed-point numbers. {\bf MP MXInt} means mixed-precision quantization using MXInt formats. Compared to MP MXInt, {\bf MP MXInt (SW-only)} does not include hardware metrics for quantization search and uses the search objective shown in Fig.~\ref{fig:search_algorithms}. {\bf MXInt8} means quantization using the MXInt with 8-bit mantissas. 

{\bf Compared with int8 and MXInt8} (in Fig.~\ref{fig:compare_mx_types}). 
The overhead in area efficiency of MP MXInt has significantly reduced compared to MXInt8 in Fig.~\ref{fig:compare_mx_types}. A major reason is the average bitwidth of MXInt mantissas has reduced to 4 bits thanks to the mixed-precision search. {\em The bitwidth reduction significantly reduces the circuit area for MXInt} while preserving the throughput, this leads to on average 1.31$\times$ area efficiency improvement. On average, MP MXInt has achieved similar area efficiency to int8. Also, the loss in accuracy caused by bitwidth reduction is negligible, where both MP MXInt and MXInt8 achieve similar accuracy compared to FP32. This demonstrates that {\em our mixed-precision quantization effectively halves the average bitwidth at no accuracy loss.}

{\bf Compared with MP int.} 
Prior work~\cite{zeng2024flightllm, qin2023fact, dettmers2022llm, kim2023squeezellm} has observed that mixed-precision quantization using fixed-point numbers can lead to efficient hardware designs with high accuracy. In our experiments, we apply fine-grained mixed-precision quantization at the tensor level for both MP int and MP MXInt. Although MP int has achieved higher area efficiency compared to int8, its accuracy loss regarding the accuracy in FP32 is significant, making MP int infeasible. This is due to the absence of dynamic ranges in fixed-point numbers, leading to significant quantization errors in deeper layers, as illustrated in Fig.~\ref{fig:motivation:dist}. Our approach, MP MXInt, preserves high accuracy with an area efficiency overhead. {\em The area efficiency difference between MP MXInt and MP int closely mirrors that between int8 and MXInt8.}

{\bf Compared with MP MXInt (SW-only).} 
A key novelty of MASE is adding hardware design metrics to quantization search, potentially leading to an efficient software and hardware co-design. Here we compare the same quantization search without hardware metrics with the MASE approach. On average, MP MXInt achieves 1.11$\times$ area efficiency of MP MXInt (SW-only). Although both approaches have achieved designs with high accuracy, {\em adding hardware metrics can guide the quantization search process towards a more area-efficient hardware design.}

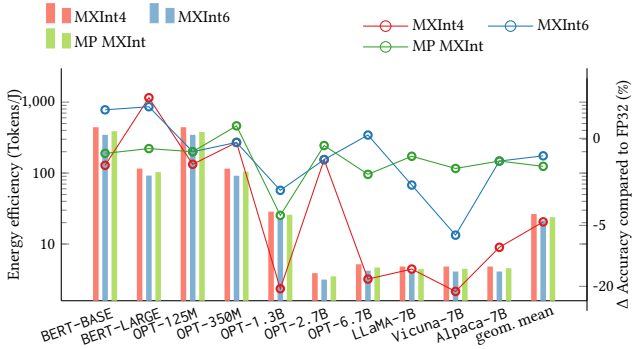
\begin{figure}[t]
\centering
\pgfplotsset{compat=1.3}
\begin{tikzpicture}[thick,scale=0.7, every node/.style={scale=0.7}]
\pgfplotsset{set layers}
\begin{axis}[
    width=0.65\textwidth,
    ybar, height=6cm, 
    ymode=log,
    legend style={at={(-0.1,1.7)},anchor=west, draw=none, fill=none}, 
    legend columns = 2,
    ylabel={Energy efficiency (Tokens/J)},
    ytick={10, 100, 1000},
    axis x line*=bottom,
    bar width=3,
    legend cell align={left},
    log ticks with fixed point,
    major x tick style = transparent,
    symbolic x coords={1, 2, 3, 4, 5, 6, 7, 8, 9, 10, 11},
    xticklabels={{\tt BERT-BASE},{\tt BERT-LARGE} ,{\tt OPT-125M} ,{\tt OPT-350M} ,{\tt OPT-1.3B} ,{\tt OPT-2.7B} ,{\tt OPT-6.7B}, {\tt LLaMA-7B}, {\tt Vicuna-7B}, {\tt Alpaca-7B}, {geom. mean}},
    xtick=data,
    ymax=3000,
    xticklabel style={xshift=-15pt, yshift=5pt, rotate=20},
    nodes near coords,
    point meta=explicit symbolic,
    nodes near coords style={
            align=center,
        },
    ]

\addplot[jcredl, draw=none, fill=jcredl] table
[x=index, y=mxint4, col sep=space] {gpu.dat};
\addlegendentry{MXInt4}

\addplot[jcbluel, draw=none, fill=jcbluel] table  
[x=index, y=mxint6, col sep=space] {gpu.dat};
\addlegendentry{MXInt6}

\addplot[jcgreenl, draw=none, fill=jcgreenl] table
[x=index, y=mixedmxint, col sep=space] {gpu.dat};
\addlegendentry{MP MXInt}

\end{axis}

\begin{axis}[
    width=0.65\textwidth,
    height=6cm, 
    axis y line*=right,
    legend style={at={(0.76,1.65)},anchor=west, draw=none, fill=none},
    legend columns = 2,
    ylabel={$\Delta$ Accuracy compared to FP32 (\%)},
    ytick={-1.707570176, -1.322219295, -0.7781512504, 0, 0.7781512504},
    yticklabels={-50, -20, -5, 0, 5},
    legend cell align={left},
    log ticks with fixed point,
    xticklabels=\empty,
    nodes near coords,
    point meta=explicit symbolic,
    nodes near coords style={
            align=center,
        },
    hide x axis,
    ]

\addplot[jcred, draw=jcred, mark=o] table
[x=index, y=mxint4, col sep=space] {gpua.dat};
\addlegendentry{MXInt4}

\addplot[jcblue, draw=jcblue, mark=o] table
[x=index, y=mxint6, col sep=space] {gpua.dat};
\addlegendentry{MXInt6}

\addplot[jcgreen, draw=jcgreen, mark=o] table
[x=index, y=mxint, col sep=space] {gpua.dat};
\addlegendentry{MP MXInt}

\end{axis}

\end{tikzpicture}
\caption{
The energy efficiency of MP MXInt sits between MXInt4 and MXInt6. MP MXInt excels in accuracy and outperforms MXInt6 by 1\% and MXInt4 by 8\% respectively on average when evaluated on the {\tt sst2} dataset.}
\label{fig:compare_mx_asic}
\end{figure}

\subsection{Evaluation across model sizes and downstream tasks}

Taking {\tt OPT} for example, we demonstrate that our approach is broadly applicable across various model sizes and tasks as shown in Fig.~\ref{fig:datasets}. Different tasks share similar dataflow hardware designs, thus we focus on model accuracy and average bitwidth. {\tt MASE IR} supports training concurrently with hardware exploration in the quantization search process. For smaller models, QAT progressively fine-tunes the model during the quantization process, achieving high accuracy; and for large models, PTQ is applied instead. Overall, Fig.~\ref{fig:datasets} agrees with the previous observations. Individual discrepancies are caused by quantization noise.

{\bf MP MXInt achieves smaller average bitwidths than MP int.}
Over all the data points, MP MXInt has smaller average bitwidths than MP int by 0.5 bit, leading to an overhead of 10\%. This overhead is due to the absence of dynamic ranges in fixed-point numbers, and more bits are required to cover the data range. Even with a larger bitwidth, MP int still fails to meet the same accuracy as MP MXInt. This indicates that the actual overhead may be larger when they have the same accuracy.

\subsection{Insights for designing future ASIC accelerators}

MASE exploits mixed-precision quantization at the tensor level to achieve high accuracy and area efficiency, leading to model-specific quantization. In applications where an accelerator may run inferences across multiple models, a more coarse-grained quantization may be amenable. Fig.~\ref{fig:compare_mx_asic} compares MP MXInt with another extreme of MXInt quantization that uniformly applies the same mantissa bits across all tensors. {\bf MXInt6} means quantization using the MXInt format with 6-bit mantissas.

{\bf Trade-off between model-specific quantization and design quality remains challenging for MX formats.} 
We evaluate the energy efficiency of the dataflow hardware accelerators, where MP MXInt sits between MXInt4 and MXInt6 due to its on average 4-bit mantissas. An interesting observation is that despite using 2 bits fewer on average, MP MXInt can still achieve better accuracy than MXInt6. This shows that {\em model-specific quantization can further push hardware efficiency significantly with no accuracy loss.} This provides insights to future accelerator design, where a trade-off needs to be explored between the granularity of model quantization and the generality of hardware designs. Such a design problem is application-specific and out of the scope of this work. However, MASE serves as a general open-source compiler and provide a platform for designers to explore potential ASIC accelerator architectures utilizing MX formats for domain-specific problems, 

\begin{table}[]
    \centering
    \caption{Runtime breakdown of the proposed toolflow, where the reported results are averaged across 10 LLMs. At the search stage, 64 trials are explored for each model.}
    \label{tab:breakdown}
{\footnotesize
\begin{tabular}{llr}
\toprule
\multicolumn{1}{c}{Stage} & \multicolumn{1}{c}{Pass name} & \multicolumn{1}{c}{Time} \\
\midrule
Pre-process & front-end & 12s \\
 & {\tt profile} & 97s \\
\midrule
\multirow{4}{*}{\begin{tabular}[c]{@{}l@{}}Search\\ (single trial)\end{tabular}} & {\tt quantize} & 5.3s \\
 & {\tt quantize (fine-tune)} & 3201s \\
 & {\tt parallelize} & 21 mins \\
 & {\tt evaluate} & 376s \\
\midrule
\multirow{2}{*}{Post-process} & {\tt emit} & 153s \\
 & {\tt synthesize} & 14.3 hours \\
 \bottomrule
\end{tabular}
}
\end{table}

\subsection{Optimization Compile Time}

Table~\ref{tab:breakdown} illustrates the runtime of MASE passes. Both the pre-process and the post-process are run once for each model in the flow, and the search process is iteratively called for a given number of trials. The front-end of MASE pre-processes the model representation when parsing from PyTorch. The search time for each step is relatively fast compared to the synthesize time, where {\em our hardware evaluation model saves significant search time by source-level hardware design analysis and avoids repeatedly calling downstream synthesis tools.}

\section{Related Work}
\label{sec:background}

In this section, we first revisit related work in quantization using block arithmetic. Then we compare MASE and {\tt MASE IR} with existing compilers and IRs. Finally, we review related work on LLM accelerator designs.

\subsection{Block Arithmetic-based Quantization}

Sharing certain components for a block of values has been widely recognized as the state-of-the-art technique for quantizing Convolutional Neural Networks (CNNs)~\cite{lin2017accurate, zhang2018lqnets}. Further explorations within this line of research have investigated grouping numbers at various granularities, including layer-wise~\cite{wu2018training}, channel-wise~\cite{krishnamoorthi2018quantizing}, and vector-wise quantization~\cite{dai2021vs}. In addition, many block floating-point variants~\cite{harma2022accuracy, dai2021vs, darvish2020pushing} have been proposed, with the core idea of grouping values into multiple blocks, and elements within each block sharing common digits. Moreover, adjusting block sizes and mantissa bitwidths across layers provides finer quantization.

There are two closest pieces of work. \cite{darvish2020pushing} proposes an approach of MXInt quantization using the same precision, while we exploit mixed-precision MXInt quantization to further push the hardware efficiency on dataflow accelerators. \cite{darvish2023shared} proposes multi-level MX formats, also known as Microscaling floating-point (MXFP), where the shared component can be non-integers, while we only restrict our scope on sharing integer components as illustrated in Fig.~\ref{fig:motivation:types}. Exploring the hardware efficiency of MXFP operators involves different challenges in both quantization search and hardware realization, which will be our future work. 

\subsection{ML Dataflow Compilers and IRs}

Most dataflow compilers for ML inference focus on DNNs. Xilinx FINN~\cite{umuroglu2017finn}, HLS4ML~\cite{fahim2021hls4ml}, DNNBuilder~\cite{zhang2018dnnbuilder}, FPGAConvNet~\cite{venieris2016fpgaconvnet}, and HIDA~\cite{ye2023hida} have shown promising results in generating efficient dataflow accelerators. However, they only support hardware mapping from quantized models using fixed-point numbers, while {\bf MASE is the first dataflow compiler that supports MX formats.} MASE comes with an open-source MX hardware operator library, and can automatically generate dataflow hardware accelerators using MX formats. Optimizations for dataflow architectures are actively studied~\cite{rucker2024revet, zhao2023sigma}, and these techniques can be orchestrated into MASE for systematic exploration with MX formats.

Most compilers for ML training and inference use software IRs, such as TorchScript~\cite{devito2022torchscript}, ONNX~\cite{bai2019onnx} and FX Torch~\cite{reed2022torch}. These IRs are often target-independent. Users need to manually add hardware intrinsic to explore target-specific optimizations, while {\tt MASE IR} targets dataflow hardware architecture with built-in hardware intrinsics. TVM~\cite{chen2018tvm} has similar IRs for GPU-specific optimizations, while MASE focuses on dataflow architectures. Languages implemented in MLIR~\cite{lattner2021mlir} or LLVM IR~\cite{lattner2004llvm} are commonly used in hardware compilers but do not support training because the back propagation functions are lost when the model is lowered from PyTorch, while {\tt MASE IR} {\bf is target-specific and keeps the model trainable for optimizations such as QAT.}

\subsection{Quantized LLM-related Accelerators}

Quantization for efficient accelerator designs has been widely studied, especially using fixed-point numbers~\cite{dettmers2022llm, frantar2022gptq, dong2019hawq, xiao2022smoothquant, yao2022zeroquant}.
Prior work focuses on custom hardware architecture for efficient inference~\cite{fan2022adaptable, ham20203, ham2021elsa, hong2022dfx, kao2023flat, li2020ftrans, lu2021sanger}. GOBO~\cite{zadeh2020gobo}, EdgeBERT~\cite{tambe2021edgebert} exploits software and hardware co-designs for accelerating transformers. FACT~\cite{qin2023fact} and FlightLLM~\cite{zeng2024flightllm} exploits mixed-precision quantization using fixed-point numbers for linear layers. 
They only focus on quantization using fixed-point numbers, and {\bf MASE is the first approach to designing LLM accelerators using mixed-precision MX quantization.}

\section{Conclusion}

LLM inference today suffers from a rapid increase of the number of parameters, leading to both memory and computing challenges. While most existing methods address these challenges by quantizing LLMs into low-precision data formats, our work highlights the ``scaling offsets'' observed in such quantization. We propose a novel dataflow compiler named MASE to explore MX formats for efficient LLM inference on dataflow hardware accelerators. {\bf MASE is the first hardware compiler to exploit hardware-aware quantization using mixed-precision MX formats.} Another contribution of MASE is that it comes with a set of open-source MX hardware operator IPs and can directly map a quantized model using MX formats into efficient dataflow hardware accelerator.

We also propose {\tt MASE IR}, an efficient software and hardware co-design IR, and show how to orchestrate existing optimizations for new data formats in {\tt MASE IR}. {\tt MASE IR} provides an open platform for designers to explore new data formats for ML hardware accelerators, minimizing their development effort and time. By exploiting mixed-precision MXInt quantization on LLMs, {\bf we verified the great potential in MXInt formats for hardware-efficient LLM inference acceleration.} Experimental data reveal that a hardware design employing mixed-precision {\tt MXInt} has achieved similar area efficiency with int8 implementation with 24\% accuracy improvements. Our results provide a performance upper bound reference for future MXInt-based accelerator designs, including ASIC accelerators.

{\bf Future ML accelerators should exploit mixed-precision MXInt formats.} Our proposed MASE compiler is the first attempt to enable exploration of future accelerators. Our future work will involve several directions. First, we plan to improve our analysis and optimization passes for deeper integration of complex MX formats, such as MXFP~\cite{darvish2023shared}. This would expand the existing MX hardware design space. Second, we plan to extend MASE to support other hardware architectures, such as systolic arrays, and explore MX formats across different granularities. This might be further extended to explore the possibility of using MASE to model and simulate ASIC MX accelerators. Finally, we will evaluate MASE on other data formats to understand the practical limitations of the approach.

\newpage

\bibliographystyle{ACM-Reference-Format}
\bibliography{ref}

\end{document}